\documentstyle[preprint,aps]{revtex}
\begin{document}
\title{}
\author{N.~Gurappa, C.~Nagaraja Kumar and Prasanta.~K.~Panigrahi 
\footnote{ email: panisp@uohyd.ernet.in}}
\address{School of Physics,University of Hyderabad, Hyderabad - 500 046 
(INDIA).\\}

\maketitle
\begin{abstract} 
We study a class of Calogero-Sutherland type one dimensional
N-body quantum mechanical systems, with potentials given by
$$ V( x_1, x_2,  \cdots x_N) = \sum_{i <j} {g \over {(x_i -
x_j)^2}} - \frac{g^{\prime}}{\sum_{i<j}(x_i - x_j)^2} + 
U(\sqrt{\sum_{i<j}(x_i - x_j)^2}),$$
where $U(\sqrt{\sum_{i<j}(x_i - x_j)^2})$'s are of specific form. It is shown
that, only for a few choices of $U$, the eigenvalue problems can be solved 
{\it exactly}, for arbitrary $g^{\prime}$. The eigen spectra of these
Hamiltonians, when $g^{\prime} \ne 0$, are non-degenerate
and the scattering phase shifts are found to be energy dependent. It is 
further pointed out that, the eigenvalue problems are amenable to solution for 
wider choices of 
$U$, if $g^{\prime}$ is conveniently fixed. These conditionally exactly 
solvable problems also do not exhibit energy degeneracy and the scattering 
phase shifts can be computed {\it only} for a specific partial wave.

\end{abstract} 
\newpage 
It has been more than twenty five years since Calogero
formulated and solved one dimensional N-body problems with
quadratic and/or inverse square, pairwise interacting potentials.$^1$
The former case, with the Hamiltonian $(\omega = 1)$
\begin{equation}
H = -\sum_{i=1}^{N} \frac{\partial^2}{\partial {x_i}^2}
+\sum_{i < j}^N
\frac{g}{(x_i -x_j)^2} + {\frac{1}{4 N}}
\sum_{i<j}^N (x_i - x_j)^2 \,\,\,,
\end{equation}
yields a degenerate bound state spectrum with 
energy given by $E_{n , k} = 2n + k + E_0$. Here, $n$,
$k$ are integers and $E_0 = \frac {1}{2} N (N - 1) (a + \frac{1}{2}) + 
\frac {1}{2} (N - 1) $
is the ground state energy; $a = \frac{1}{2} \sqrt{1 + 2g}$.  
The purely inverse square potential gives rise to only scattering states, 
where the phase shifts are independent of energy.
Although, this model was projected as a toy model, pending
extension to higher dimensions, it has since been found to
be of relevance in many other contexts e.g., spin systems,$^2$
fractional statistics,$^3$ conformal field theory and chaos.$^4$
Soon after Calogero's work, Sutherland extended this to a model,
where interaction takes place on a circle.$^5$ Recently, another
N-body problem with the potential,$^6$
\begin{equation} 
V( x_1, x_2, x_3 \cdots x_N) = \sum_{i <j}
{g \over {(x_i - x_j)^2}} - \frac{g^{\prime\prime}}{\sqrt{\sum_{i<j}(x_i -
x_j)^2}} \, \, \, \, ,
\end{equation}
\noindent 
has been shown to be exactly solvable with  degenerate spectrum:
$E_{n , k} = - {g^{{\prime\prime}2} \over {4 N}} {1 \over 
{(n + k + b + \frac {1}{2})^2}},$ where $b = \frac{1}{2}(N - 3) + 
\frac{1}{2}N(N - 1) (a + \frac{1}{2})$.
In addition to the pairwise `centrifugal' term, this
potential contains a N-body interaction term. It is then
natural to ask, if there exists other potentials for which the
N-body problems can be solved exactly. This is of importance, since
exactly solvable N-particle systems with long-range interactions are few and 
far between. It is also of interest to find out the common origin of 
solvability of these models. The study of the degeneracy structure  
is also of significance because of its deep connection with symmetry.

In this letter, we present a class of potentials, containing both
pairwise inverse square and N-body interaction terms, of the type
\begin{equation}
 V( x_1, x_2, x_3
\cdots x_N) = \sum_{i <j} {g \over {(x_i - x_j)^2}} -
\frac{g^{\prime}}{\sum_{i<j}(x_i - x_j)^2} + U (\sqrt{\sum_{i<j}(x_i - 
x_j)^2})\,\,
\,.
\end{equation}

It is shown that, when $U = {\frac{1}{4 N}}
\sum_{i<j}^N (x_i - x_j)^2$ and
$g^{\prime} \ne 0$, the above generalized Calogero type model
can also be solved {\it exactly} with $E_{n,k} = 2 n + \lambda_k + \frac{3}{2}$.
Here, $\lambda_k = -{1\over 2} + \sqrt{(k + b)^2 - {{g^{\prime}} \over {N}}}$.
Similarly, when $U = - \frac{g^{\prime\prime}}{\sqrt{\sum_{i<j}(x_i - x_j)^2}}$,
the above model generalizes that of Khare, with $E_{n , k} = 
- {g^{{\prime\prime}2} \over {4 N}} {1 \over {(n + \lambda_k + 1)^2}}$; 
$\lambda_k$ is as given above. The striking feature 
is that, these generalized models do not exhibit energy level degeneracy.
In the corresponding scattering problems, the phase shifts are found to be
energy dependent 
unlike the example of the Calogero model. For arbitrary $g^{\prime}$, we have
not found it 
possible to solve the eigenvalue problem for other many-body potentials.

However, we find that, for some other choices of $U$, the eigenvalue 
spectra can also
be obtained exactly, provided the parameter $g^{\prime}$ takes suitable 
values. 
These models provide the N-particle analogs of the conditionally exactly 
solvable (CES) systems in one variable quantum mechanics.$^7$ The scattering 
phase shifts for these CES examples can be computed only for a given
partial wave in the $k$ channel and is found to be energy dependent.

The N-body Hamiltonian of interest describes identical particles
of mass $m$ in one dimension  $ (\hbar = 2m = 1; \,\, g >
-\frac{1}{2}) $
\begin{equation}
H= -\sum_{i=1}^{N} \frac{\partial^2}{\partial {x_i}^2}
+\sum_{i < j}^N
\frac{g}{(x_i -x_j)^2} - \frac{g^{\prime}}{\sum_{i<j}(x_i - x_j)^2}
+ U(\sqrt{\sum_{i<j}(x_i - x_j)^2})\,\,\,,
\label{3}
\end{equation}
where, we have to solve the eigenvalue equation 
\begin{equation}
H \psi = E \psi \, \, \, \, \, .
\label{4}
\end{equation}
Here, $\psi$ is taken to be a translation invariant eigenfunction. 
We restrict our attention to the sector of the configuration
space with a definite ordering of the particles, 
\begin{equation}
{\it i.e.}, \,\,\,\,x_i \ge x_{i+1} \, \, \, \, .
\end{equation}
This can be done in one dimension, since the particles cannot overtake each 
other because of the mutual repulsive potential. The normalizable solutions of 
Eq. (5) can be written as
\begin{equation}
\psi(x) = Z^{a +\frac{1}{2}}\,\,\phi(r)\,\, P_k (x) \,\,\,,
\label{6}
\end{equation}
where, $Z = \prod_{i<j}^N  (x_i -x_j), \,\,\,\, i= 1,2,\cdots N,$ \,\,
and $ a = \frac{1}{2} \sqrt{1+2 g}$. Here, $r = \sqrt{{\frac{1}{N}}
\sum_{i<j}^N (x_i - x_j)^2}$ 
and $P_k(x)$ is a homogeneous, translation invariant, symmetric polynomial of 
degree $k$ in $x_i$'s, satisfying the generalized Laplace equation
\begin{equation}
\left[ \sum_{i=1}^{N} \frac{\partial^2}{\partial {x_i}^2} + 2 (
a + \frac {1}{2}) \sum_{i<j}^N \frac{1}{(x_i - x_j)}
(\frac{\partial}{\partial x_i} -\frac{\partial}{\partial x_j}) \right]
P_k (x) = 0 \,\,\,.
\label{7}
\end{equation}
On substituting Eq. (7) in Eq. (5) and making use of Eq. (8), along with the 
properties of the translation invariance and homogeneity of  $P_k 
(x)$,$^1$ one 
finds that $\phi(r)$ satisfies the following differential equation 
\begin{equation}
-[\phi^{\prime\prime} (r) + \{ 2 k + 2 b + 1 \} \frac{1}{r}
\phi^{\prime} (r) ] -[ \frac{g^{\prime}}{N r^2} - U(\sqrt{N}r) + E ] \phi (r) 
=0 
\,\,\,,
\label{8}
\end{equation}
where $ b= \frac{N(N-1)}{2} a +\frac{N(N+1)}{4} - \frac{3}{2}$. \\ 
Removing the first derivative term in Eq. (9) by writing,
\begin{equation}
\phi(r) = r^{-( l +1)} \chi (r) \,\,\,,
\end{equation} 
we find  $\chi(r)$ satisfies the radial Schr\"odinger equation
\begin{equation}
-\chi^{\prime\prime}(r) + [ \frac{l(l+1)}{r^2} - \frac{g^{\prime}}{N r^2}
+ U(\sqrt{N}r) ] \chi (r) = E \chi(r)\,\,\,,
\end{equation}
with the identification of $ l =   k +  b -\frac{1}{2}$.

When $U(r)=0$, this problem is similar to that of the Calogero problem
with purely inverse square interaction
and if $U(r)$ is chosen to be the harmonic potential, the problem reduces 
to that of the Calogero type with harmonic confinement. Similarly, if $U(r)$ 
is taken to be of Coulomb form, the problem generalizes the one given in 
Ref. 6. These two models can be solved for general values of $g^{\prime}$, 
provided, it is suitably chosen so that the quantum mechanical problem
is well defined in the variable `$r$'.$^8$ In both these examples 
however, the degeneracy is completely removed unlike the $g^{\prime} = 0$ 
case.

The energy eigenfunctions and eigenvalues for the harmonic case 
($\omega =  1$), {\it i.e.,} when $U = {{\frac{1}{4N}}\sum_{i<j}^N (x_i - 
x_j)^2}$ reads, 
\begin{equation}
\psi_{n,{\lambda_k}} =  Z^{a +\frac{1}{2}}\,\,\
 y^{({\lambda_k} - l) \over 2} e^{-{1\over 2} y} 
{L_n}^{{\lambda_k} + {1\over 2}} (y) \,\, P_k (x) 
\end{equation}
and
\begin{equation}
E_{n,k} = 2 n + {\lambda_k} + \frac {3}{2}\,\,\,.
\end{equation}
Here $ y = {1\over 2} r^2 $ and
\begin{equation}
\lambda_k =  -{1\over 2} + \sqrt{(k + b)^2 - {{g^{\prime}} \over {N}}}\,\,\,.
\end{equation}
Notice that the energy levels are non-degenerate for finite values of 
$g^{\prime}$ and $N$. In the limit $g^{\prime} \rightarrow 0$ , the above 
result smoothly goes over to the Calogero case. This is analogous to the three
dimensional oscillator problem, where the presence of an additional 
centrifugal
term removes the degeneracy.$^{9}$

When $ U(r) = - {{g^{\prime\prime}} \over 
{\sqrt{\sum_{i<j}^N (x_i - x_j)^2}}}$ {\it i.e.,} Coulomb type
and $g^{\prime} \ne 0$, the corresponding eigenvalue 
problem can be solved analogous to the generalized Calogero system 
and the eigenvalues and the corresponding eigenfunctions are given by 
$$E_{n , k} = - {g^{{\prime\prime}2} \over {4 N}} {1 \over 
{(n + \lambda_k + 1)^2}}$$
and
$$\psi_{n,{\lambda_k}} =  Z^{a +\frac{1}{2}}\,\,\
y^{({\lambda_k} - l)}\,\,\,  e^{-{1\over 2} y} \,\,\,
{L_n}^{2 {\lambda_k} + 1} (y) \,\, P_k (x) \,\,\,.$$
respectively.
Here $ y = {{g^{{\prime\prime}2}  r } \over {2 N (n + \lambda_k + 1)^2}}$;
$\lambda_k$ is given by Eq. (13) and $l = k + b - \frac{1}{2}$.
One can immediately see that the energy levels are non-degenerate. 

In order to discuss the scattering scenario, one needs to find the positive
energy solutions {\it i.e.,} $E = p^2  > 0$ and their asymptotic 
behaviour, when all particles are far apart from each other. For the sake of
comparison, we briefly outline below the derivation of the scattering phase 
shifts in the Calogero model. In this case, the positive energy eigenfuctions 
are given by (using the same notations as in Ref. 1) 
\begin{equation}
\psi_k = Z^{a+\frac {1}{2}} r^{-(A + k)} J_{(A + k)} P_k(x)\,\,\,. 
\end{equation}
The most general stationary eigenfuction can be written as
\begin{equation}
\psi = Z^{a+\frac {1}{2}} \sum_{k=0}^{\infty} \sum_{q=1}^{g(N,k)} c_{kq} 
r^{- A - k} J_{A + k} P_{k q}(x)\,\,\,,
\end{equation}
where, $J_{b}$ is the Bessel function; $A = \frac{1}{2}(N - 3) + 
\frac{1}{2} N (N - 1) (a + \frac{1}{2})$ and $p = \sqrt {|E|}$.
The asymptotic limit of Eq. (16) is given by
\begin{equation}
\psi \sim \psi^{\rm in} + \psi^{\rm out}\,\,\,.
\end{equation}
Here,
$$\psi_{\rm in} \equiv {(\frac{1}{2} \pi p r )^{-\frac{1}{2}}} Z^{a + \frac{1}
{2}}r^{-A} \sum_{k=0}^{\infty} \sum_{q=1}^{g(N,k)} c_{kq} 
e^{i(A+k+\frac{1}{2})
\frac{1}{2} \pi - i p r } P_{kq}  (x)$$
and
$$\psi_{\rm out} \equiv {(\frac{1}{2} \pi p r )^{-\frac{1}{2}}} Z^{a + 
\frac{1}
{2}}r^{-A} \sum_{k=0}^{\infty} \sum_{q=1}^{g(N,k)} c_{kq} e^{-i(A+k+\frac{1}
{2})\frac{1}{2} \pi + i p r } P_{kq}  (x) \,\,\,.$$
This can also be characterized as an incoming and outgoing plane wave, 
involving $N$-particles as
\begin{equation}
\psi_{\rm in}  \equiv c \exp  \{ i \sum_{i=1}^N p_i x_i \}\,\,
\end{equation}
and
\begin{equation}
\psi_{\rm out}  \equiv e^{-i A \pi} c \exp  \{ i \sum_{i=1}^N p_{N + 1 - i}
 x_i \}\,\,,
\end{equation}
respectively, by choosing $c_{kq}$'s appropriately and making use of the 
properties of symmetry and the homogeneity of $P_{kq} (x) $. Notice that the 
final momenta, $p_i^{\prime}$, is given by $p_i^{\prime} = p_{N + 1 - i}$. One
can also immediately see that, the scattering phase shifts are energy 
independent.

Following the Calogero case, in our model with $g^{\prime} \ne 0$
and in the absence of harmonic confinement, the positive energy eigenfunctions
are found to be 
\begin{equation}
\psi_k = Z^{a+\frac {1}{2}} r^{-(k+b)} J_{{\nu}_k} P_k(x) \,\,\,.
\end{equation}
Here, $J_{\nu_k}$ is the Bessel function;
${\nu_k} =  \sqrt {(k + b)^2 - \frac{g^{\prime}}{N}}$ and $p = \sqrt {|E|}$.
In the asymptotic limit, $\psi_k$ can be written as
$$\psi_k \sim \psi_k^{in} + \psi_k^{out}$$
where
\begin{equation}
\psi_k ^{in} \sim (2 \pi p r)^{-\frac{1}{2}} Z^{a+\frac 1 2} 
r^{-(k+b)} e^ {i({\nu_k} + \frac {1}{2}){\frac {\pi}{2}} - i p r} 
P_k(x) \nonumber
\end{equation}
and
\begin{equation}
\psi_k ^{out} \sim (2 \pi p r)^{-\frac{1}{2}} Z^{a+\frac 1 2} 
r^{-(k+b)} e^ {-i({\nu_k} + \frac {1}{2}){\frac {\pi}{2}} + i p r} 
P_k(x) \,\,\,.
\end{equation}

However, the phase shift for each partial wave in the $k$ channel
is $e^{-i \pi {\nu_k}}$, which is energy 
dependent. Although, we can write the most general stationary eigenfunction as
in Eq. (16), the initial scattering 
situation characterized by the initial momenta $p_i\,\,\,(i = 1,2,..., N)$ can 
not go over to a final configuration characterized by the final momenta 
$p_i^{\prime} = p_{N + 1 - i}$ unlike in the
Calegero model. This is because, all the partial waves with the quantum number
$k$ get phase shifted, with phase shifts depending on energy.

Apart from these  cases, we observe that, in the radial
variable `$r$', the reduced Schr\"odinger equation is not
amenable to solution for other cases in the presence of
the centrifugal term. At this stage, if one imposes 
the condition that $g^{\prime} = N l (l+1)$, akin to the CES quantal
problems,$^7$ the corresponding Schr\"odinger equation reduces to
\begin{equation} 
-\chi^{\prime\prime}(r)  + U ( \sqrt{N} r ) \,\,\,\chi(r) = E \,\,\,
\chi(r)\,\,\,.
\label{11}
\end{equation}
Here, one observes that $U(\sqrt{N} r) $ can be
chosen to be any of the known exactly / quasi-exactly solvable
potentials,$^{10}$ in which case the energy spectra of the full Hamiltonian 
can be
obtained by using the known results.$^{11}$ In these cases, the eigenfunctions 
of the original $N$ body problem can be written as
\begin{equation}
\psi_n (x) = Z^{a +\frac{1}{2}}\,\,\,r^{-(l+1)}\,\,\, \chi_n(r)\,\,\, 
P_k (x) \,\,\,,
\label{11a}
\end{equation} 
which is same as Eq. (7) but with a fixed $l$ (hence $k$). 

In this case, it is interesting to note that, once  $k$ is
fixed, the energy spectrum has no degeneracy. We observe that
this property will be a generic feature for all these
CES N-body problems.

For the sake of illustration, $U(\sqrt{N} r)$ is chosen as the Rosen-Morse 
potential.$^{11}$ The potential reads,
\begin{equation}
U(r) = ( {A^2} + {B^2} + A \beta \sqrt{N} )\,{\rm cosech}^2 (\beta 
\sqrt{N}
 r ) - B ( 2A + \beta \sqrt{N} )\,\coth(\beta \sqrt{N} r)\,{\rm 
cosech}
(\beta \sqrt{N} r)\,.
\end{equation}
The eigenvalues and the  eigenfunctions are       
\begin{equation}
E_n= -(A - {n}{\beta \sqrt{N}})^2 
\end{equation}
and
\begin{equation}
\psi (x) = Z^{a +\frac{1}{2}}\,\,\,r^{-(l + 1)}\,\,\,
( y - 1)^{(\lambda - s)\over 2}\,\,\,( y + 1)^{-(\lambda + s)\over 2}\,\,\,
P_n^{(\lambda - s -  \frac{1}{2}) , (-\lambda - s -\frac{1}{2})} (y) 
\,\,\, P_k (x) 
\end{equation}
respectively. Here $P_n^{\mu, \nu} (y)$'s are the Jacobi polynomials,
$y= \cosh \beta \sqrt{N} r $, $ \lambda = \frac{B}{\beta \sqrt{N}}$ ,
$ s = \frac{A}{\beta \sqrt{N}}$,
and $ l = k + b - \frac{1}{2}$ 

The scattering solutions are obtained by analytically 
continuing $n$ to ($s-\frac{ip}{\beta \sqrt{N}}$), as 
\begin{eqnarray}
\chi = ( y - 1)^{(\lambda - s)\over 2}\,\,\,& ( y + 1)^{-(\lambda + s)\over 2} 
\,\,\,\frac{\Gamma(\beta \sqrt{N} + 1 + s - \frac{ip}{\beta \sqrt{N}})}
{ {n}! \Gamma (\beta \sqrt{N} + 1 )} \times \,\,\nonumber\\
& F(\frac{ip}{\beta \sqrt{N}} -s, -\frac{ip}{\beta \sqrt{N}} -s ; \lambda - 
s +\frac {1} {2};  \frac{1-y}{2} ),
\end{eqnarray}
where $F(y)$ is a hypergeometric function. Taking the asymptotic limit of 
$\psi$, one gets 
\begin{eqnarray}
\psi_k &\sim & Z^{a+\frac {1}{2}} \,\,\,r^{-(k+b +\frac 1 2)} \,\,\,P_k(x) 
\,\,\,
\left[e^{-ipr} + e^{i\delta } \,\,\,e^{ipr} \right] \nonumber\\
& = & \psi_k^{\rm in} + \psi_k^{\rm out}
\end{eqnarray} 
where 
\begin{equation}
e^{i\delta} = \frac{2^{-4 i q} \Gamma (2 i q ) \Gamma( -s -i q)
\Gamma(\lambda +\frac{1}{2} - i q )}{\Gamma ( - 2 i q )\Gamma(- s + i q )
\Gamma(\lambda + \frac{1}{2} + i q ) }\,\,\,.
\end{equation}
Here $ q = \frac{p}{\beta \sqrt{N}}$ and $\delta$ is the scattering phase 
shift. It is worth noticing that, in this CES potential, 
$\frac{g^{\prime}}{r^2}$
term is $k$ dependent and as a result of which the degeneracy in $k$
has been lifted. Although the incoming wave $\psi_{\rm in} (x)$ can be
decomposed into partial waves as in the Calogero case, the phase shift can be
obtained only for the specific value of $k$ that appears in the coupling 
parameter $g^{\prime}$. One cannot obtain any further information 
regarding the scattering of other partial waves unlike the previously
considered examples.

To conclude, we have obtained the complete energy spectra for new exactly
solvable (ES) as well as a wide class of CES 
N-body problems in one dimension. In fact, addition of the term
`$ - \frac{g^{\prime}}{\sum_{i<j}(x_i - x_j)^2} $' in the potential 
enabled us to generalize all the known ES, CES and QES
one variable quantum mechanical problems into 
ES or partially solvable $N$-body systems in one dimension.
The energy non-degeneracy is found to be a generic feature of all the above 
models, except in the case when $U(r)$ is of harmonic or Coulomb form with
$g^{\prime} = 0$ {\i.e.,}Calogero or Khare case. In ES case, the 
energy dependent scattering phase shifts are obtained for each partial wave,
whereas the phase shift in the CES case is known for 
only one partial wave.

We also remark that, studying the algebraic structure, as was done
in the Calogero model,$^{12}$ is of great relevance in order to 
find the underlying symmetries, if any, for the present models. Construction 
and analysis of the coherent states$^{13}$ is another direction worth 
exploring, since they will provide a better understanding of the 
semi-classical behaviour of these quantum mechanical models. 

We acknowledge valuable discussions with Profs. A. Khare, V. Srinivasan and
S. Chaturvedi. We also thank R.S. Bhalla, E. Hari Kumar and G.V. Rao for
useful comments. N.G would like to thank U.G.C (India) and
C.N.K to C.S.I.R (India) for financial support through JRF and RA schemes
respectively.

\noindent{\bf References}
\begin{enumerate}
\item F. Calogero, {\it Jour. Math. Phys.} {\bf 10}, 2191 (1969); ibid
{\bf 12}, 419 (1971).
\item F.D.M. Haldane, {\it Phys. Rev. Lett.} {\bf 60}, 635 (1988); 
B.S. Shastry, ibid {\bf 60}, 639 (1988).
\item F.D.M. Haldane, {\it Phys. Rev. Lett.} {\bf 67}, 937 (1991);
 M.V.N. Murthy and R. Shankar, ibid {\bf 73}, 3331 (1994).
\item B.D. Simons, P.A. Lee and B.L. Altshuler, {\it Phys. Rev. Lett.} 
{\bf 72}, 64 (1994).
\item B. Sutherland, {\it Jour. Math. Phys.} {\bf 12}, 246 (1971).
\item A. Khare, preprint hep-th 9510096, to appear in {\it 
      J. Phys}. {\bf A}.
\item A. de Souza Dutra, {\it Phys. Rev.} {\bf 47 A}, R2435 (1993).
\item A. Gangopadhyaya, P.K. Panigrahi and U. Sukhatme, 
      {\it Jour. Phys.} {\bf A 27}, 4295 (1994).
\item L. Landau and E.M. Lifshitz, {\it  Quantum Mechanics: Non-relativistic 
      theory}, Oxford Univ. Press (1975).
\item For a review of QES systems, the readers are referred to 
      M.A. Shifman, {\it Int. Jour. Mod. Phys.} {\bf A 4}, 2897 (1989).
\item F. Cooper, A. Khare and U. Sukhatme, {\it Phys. Rep.} {\bf 251}, 267
      (1995).
\item M.A. Olshanetsky and A.M. Perelomov, {\it Phys. Rep.} {\bf 71}, 314 
     (1981); {\bf 94}, 6 (1983);
     A.P. Polychronakos, {\it Phys. Rev. Lett.} {\bf 69}, 703 (1992); L. Brink,
     T.H. Hansson, S.E. Konstein and M. Vasiliev, {\it Nucl. Phys.} {\bf B 384},
     591 (1993); H. Ujino and M. Wadati, {\it Jour. Phys. Soc. Jpn} {\bf 64},
     39 (1995); V. Narayanan and M. Sivakumar, preprint hep-th/9510239;
     N. Gurappa and P.K. Panigrahi, preprint cond-mat/9602059; to appear 
     in {\it  Mod. Phys. Lett.} {\bf A}.
\item G.S. Agarwal and S. Chaturvedi, {\it J. Phys.} {\bf A 28}, 5747 (1995).
\end{enumerate}
\end{document}